\documentclass[aps,prl,preprint,groupedaddress]{revtex4}

\usepackage{graphicx}

\usepackage{amsmath,amsthm,amsfonts, latexsym}

\newcommand{\mtx}[2]{\left(\begin{array}{#1}#2\end{array}\right)}

\begin{document}


\title{Entanglement Sharing in Real-Vector-Space Quantum Theory}


\author{William K. Wootters}
\affiliation{Department of Physics, Williams College, Williamstown, MA 01267, USA
\\
Department of Applied Physics, Kigali Institute of Science and Technology,
B.P.~3900, Kigali, Rwanda}


\date{\today}

\begin{abstract}
The limitation on the sharing of entanglement is a basic feature of quantum theory.  For example, if two qubits are completely entangled with each other, neither of them can be at all entangled with any other
object.  In this paper we show, at least for a certain standard definition of entanglement, that this feature is lost when one replaces the usual complex vector space of quantum states
with a real vector space.  Moreover, the difference between the two theories is extreme: in the real-vector-space theory,
there exist states of arbitrarily many binary objects, ``rebits,'' in which every rebit in the system is maximally entangled
with each of the other rebits.  
\end{abstract}

\pacs{}

\maketitle


\section{Introduction}

It is a pleasure to offer this paper in celebration of the birthdays of Daniel Greenberger and Helmut Rauch, 
both of whom have contributed immensely to our understanding of quantum mechanics.  The particular aspect
of quantum mechanics that I address in this paper is the sharing of entanglement among more than two 
particles, a subject that was opened up many years ago by the Greenberger-Horne-Zeilinger theorem \cite{GHZ}. 

A pure quantum state of two objects is called entangled if is not a product state.  A mixed state is called entangled if it cannot be
written as a mixture of product states.  In terms of quantitative measures based on these definitions,  it is well known that 
quantum theory strongly limits the sharing of entanglement. 
For example, if two qubits are maximally entangled with each other---they could be in the state
$(1/\sqrt{2})(|00\rangle + |11\rangle)$---then neither of them can be at all entangled
with a third object.  This fact is sometimes called the ``monogamy'' of entanglement, and it is
one of the features that distinguishes entanglement from classical correlation.  In the classical case, it is easy to construct a probability
distribution in which, for example, $n$ coins are all pairwise maximally correlated: the distribution might assign probability
one-half to each of the configurations ``all heads'' and ``all tails" \footnote{By ``maximal correlation'' between two
coins, I mean maximal mutual information.  That is, each coin is equally likely to be heads or tails, but
the state of one coin can be inferred perfectly from the state of the other.}.  But no such situation is possible with quantum entanglement.

Even when entanglement is not maximal, quantum theory imposes restrictions.  Suppose qubits $B_1, B_2, \ldots, B_n$
are all partially entangled with qubit $A$.  The entanglement between $A$ and $B_j$ can be quantified by the 
concurrence $C_{AB_j}$ \cite{Hill, Wootters}, which is computed from the density matrix of the pair $AB_j$.  (See the following section for 
the definition.)  One can show that the various concurrences with qubit $A$ are restricted by the inequality \cite{Coffman, Osborne}
\begin{equation}
C_{AB_1}^2 + \cdots + C_{AB_n }^2 \le 1. \label{OV}
\end{equation}
In this sense qubit $A$ has only a limited amount of entanglement to share, and any entanglement that it has with 
qubit $B_1$ reduces the amount available for qubits $B_2$ through $B_n$ \footnote{Similar limitations apply
to nonlocal quantum correlations.  For a review see Ref.~\cite{Zeevinck}}.  The inequality (\ref{OV}) has been used in the
analysis of entanglement in many-body systems, especially in the study of quantum phase transitions.  (For a review
of this and other applications of entanglement theory to many-body systems, see Ref.~\cite{Amico}.)

To get a better understanding of any feature of quantum mechanics, it is often useful to ask whether that feature
appears in other, hypothetical theories that are similar to quantum mechanics.   One might be inclined to think that the limitation on entanglement sharing is so fundamental that it would be characteristic of
a wide class of such theories.  In particular, one might expect this feature to appear
in real-vector-space quantum theory, a theory that is essentially identical to ordinary quantum theory except that the usual complex vector space of pure states is replaced by a 
real vector space. In this paper we consider a collection of binary real-vector-space objects---``rebits''---and ask whether there is a limit on their sharing of entanglement.  We find that if we again use concurrence to quantify entanglement, there is in fact no such limit.  Specifically, for a system of arbitrarily many rebits, there exist states in which each rebit in the system is {\em maximally} entangled with each of the other rebits.  Thus the limitation on entanglement sharing, unlike many other quantum features such as the no-cloning principle and the violation of Bell's inequality, depends on the field over which the vector space of states is defined. 

We begin by recalling how entanglement can be quantified.

\section{Quantifying entanglement}

Much work has been done on the quantification of entanglement, and many different entanglement measures have appeared in the literature (for reviews see Refs.~\cite{Plenio, Zyckowski, Horodecki}).  The measures differ from each other partly with respect to their purpose.  One can, for example, quantify the entanglement of a bipartite mixed state according to the amount of pure entanglement one can extract from the state.  Here I focus on the opposite question: what resources are required to {\em create} a given state?  

The standard unit of entanglement is the ebit, which is the amount of entanglement in the two-qubit state $|\Phi^+\rangle = (1/\sqrt{2})(|00\rangle + |11\rangle)$.  This state is as fully entangled as a two-qubit state can be.  If two spatially separated participants, each holding one qubit and initially sharing no entanglement, want to get their qubits into the state $|\Phi^+\rangle$, one of the participants will have to transmit a qubit's worth of quantum information to the other; the state cannot be created with classical communication alone. We note that the closely related pure state $|\Phi^-\rangle = (1/\sqrt{2})(|00\rangle - |11\rangle)$, which we will invoke shortly, is likewise maximally entangled.  

We can quantify the entanglement of a {\em mixed} state by considering the pure states involved in a decomposition of the state \cite{BDSW, Popescu}.  Consider, for example, a density matrix $\omega$ that 
is an equal mixture of $|\Phi^+\rangle$ and $|\Phi^-\rangle$:
\begin{equation}
\omega = (1/2)(|\Phi^+\rangle\langle \Phi^+| + |\Phi^-\rangle\langle \Phi^-|).
\end{equation}
We could create the state $\omega$ by randomly choosing to create either $|\Phi^+\rangle$ or $|\Phi^-\rangle$ and then
discarding the information that tells us which state was created.  This method of constructing $\omega$ costs one ebit, in the sense that in 
either case, one has to create
a pure state having one ebit of entanglement.  But a given mixed state can be written in more than one way as a mixture of pure states.  (In fact there is always a continuum of possible decompositions.)  It is a simple mathematical fact that $\omega$ can also be written as
\begin{equation}
\omega = (1/2)(|00\rangle\langle 00| + |11\rangle\langle 11|).
\end{equation}
Thus, rather than using the entangled states $|\Phi^+\rangle$ and $|\Phi^-\rangle$, one could construct the state $\omega$ by randomly choosing to create either $|00\rangle$ or $|11\rangle$ and then discarding the record.  This latter method costs zero ebits, because both $|00\rangle$ and $|11\rangle$ are product states that can be created locally; no quantum communication is required.  We therefore say that the state $\omega$ is unentangled: even though it {\em could} be created in a difficult way that involves entangled states, those states 
are not required for its construction, and the state can in fact be generated locally.  

Of course we also want to quantify the entanglement of {\em non-maximally} entangled pure states of two qubits, and of mixed states constructed from such pure states.  We begin by noting that any pure two-qubit state can, by local unitary transformations, be brought to the form $|\psi\rangle = \alpha |00\rangle + \beta |11\rangle$, where $\alpha$ and $\beta$ are real and non-negative and $\alpha^2 + \beta^2 = 1$.  The standard measure of entanglement for such a state is given by the entropic formula $E(|\psi\rangle) = -(\alpha^2 \log \alpha^2 + \beta^2 \log \beta^2)$.  This is the formula to use if one wants to measure the entanglement in ebits, because it gives the asymptotic conversion rate between the state $|\psi\rangle$ and a maximally entangled state \cite{compression}.  But any monotonic function of $E$ can serve as an alternative measure (though it may lack this direct interpretation in terms of asymptotic conversion).  In this paper it is convenient to use the simpler quantity $C(|\psi\rangle) = 2\alpha\beta$, which is called the concurrence of the state $|\psi\rangle$ \cite{Hill, Wootters}.  Note that $C$ ranges from 0 to 1.  

To define the concurrence of a {\em mixed} state of two qubits, we use the ideas illustrated above in our discussion of the state $\omega$.  As we have noted, a given mixed state $\rho$ can be written in infinitely many ways as a mixture of pure states.  The concurrence of 
$\rho$ is defined to be the average concurrence of the pure states in a decomposition of $\rho$, minimized over all possible 
decompositions.  That is,
\begin{equation}
C(\rho) = \inf \left( \sum p_j C(|\phi_j\rangle) \right), \hspace{2mm} \hbox{where} \hspace{2mm} \sum p_j |\phi_j\rangle\langle \phi_j|
= \rho.
\end{equation}
It happens that the minimum can be evaluated in general to get an analytic formula for $C(\rho)$ \cite{Hill, Wootters}.  We will not need the formula in this paper, but 
I write it here because it will be interesting to compare it shortly to the analogous formula for the real-vector-space
theory.
\begin{equation}
C(\rho)  = \max \{ \lambda_1 - \lambda_2 - \lambda_3 - \lambda_4, 0 \},  \label{conc}
\end{equation}
where the $\lambda_i$'s are the square roots of the eigenvalues
of $\rho (\sigma_y \otimes \sigma_y) \rho^* (\sigma_y \otimes \sigma_y)$, ordered so that $\lambda_1$ is the largest.  Here $\sigma_y$ is the Pauli matrix $\mtx{cc}{0 & -i \\ i & 0}$ and the asterisk
indicates complex conjugation in the standard basis.  

As an interesting example, consider the mixed state $\omega'$ which, like the state $\omega$ considered above, is an equal mixture of two maximally entangled
pure states:
\begin{equation}
\omega' = (1/2)(|\Phi^-\rangle \langle \Phi^-| + |\Psi^+\rangle \langle \Psi^+|),
\end{equation}
where $|\Psi^+\rangle = (1/\sqrt{2})(|01\rangle + |10\rangle)$.  It turns out that, like $\omega$, the state $\omega'$ can be written
as a mixture of product states and therefore has zero concurrence.  This fact can be verified by computing the 
concurrence using Eq.~(\ref{conc}), but it is simpler to exhibit a specific decomposition of $\omega'$ into 
product states: 
\begin{equation}
\omega' = (1/2)(|\xi^+\rangle\langle \xi^+| + |\xi^-\rangle\langle \xi^-|),
\end{equation}
where 
\begin{equation}
|\xi^+\rangle = \frac{1}{\sqrt{2}}(|\Phi^-\rangle + i|\Psi^+\rangle) = \frac{1}{2}(|0\rangle + i|1\rangle)\otimes(|0\rangle +i |1\rangle) 
\end{equation}
and
\begin{equation}
|\xi^-\rangle = \frac{1}{\sqrt{2}}(|\Phi^-\rangle - i|\Psi^+\rangle)= \frac{1}{2}(|0\rangle - i|1\rangle)\otimes(|0\rangle -i |1\rangle).
\end{equation}
These last two equations show that the pure states of this decomposition are indeed unentangled.  
It turns out that the imaginary unit $i$ is absolutely necessary to achieve such a product-state decomposition of $\omega'$.  The impossibility of this sort of decomposition in the real-vector-space theory is what will lead to the very different entanglement sharing properties of that theory, as we will see.  

We quantify entanglement in the real-vector-space theory just as we did in the complex case.  A general pure state
of two rebits can, by local orthogonal transformations, be brought to the form $\alpha|00\rangle + \beta|11\rangle$ where $\alpha$ and $\beta$ are non-negative.  And as before, we define the concurrence to be $C = 2\alpha\beta$.  The concurrence of a mixed
state is again defined to be the minimum, over all pure-state decompositions, of the average concurrence of the component states \cite{Fuchs}.  That is,
\begin{equation}
C_R(\rho) = \inf \left( \sum p_j C(|\phi_j\rangle) \right), \hspace{2mm} \hbox{where} \hspace{2mm} \sum p_j |\phi_j\rangle\langle \phi_j|
= \rho.
\end{equation}
This definition looks exactly the same as in the complex case, but it is different: the states
$|\phi_j\rangle$ appearing here are required to be {\em real} state vectors.  As a consequence
of this restriction, the real-vector-space concurrence $C_R(\rho)$ of a given real density matrix $\rho$ does not have to have 
the same value as the usual concurrence $C(\rho)$ of the same density matrix.

As in the complex case, the minimization required by the definition of concurrence can be performed in general, so that one can again write down an analytic formula, as shown by Caves, Fuchs, and Rungta \cite{Fuchs}.  But the formula is different.  It is simply
\begin{equation}
C_R(\rho) =\big| \hbox{Tr}\,[(\sigma_y \otimes \sigma_y) \rho] \big|.  \label{realconc}
\end{equation}
Thus the concurrence of a pair of rebits is particularly easy to compute.  Rather than having to find the eigenvalues of
a matrix as in Eq.~(\ref{conc}), one simply has to compute the average value of a particular operator.  

Let us apply this formula to the state $\omega' = (1/2)(|\Phi^-\rangle\langle\Phi^-| + |\Psi^+\rangle\langle \Psi^+|)$ that we have been discussing.  One can rewrite this state as $
\omega' = (1/4)(I \otimes I + \sigma_y \otimes \sigma_y)$,
where $I$ is the $2 \times 2$ identity matrix.  We then have that 
\begin{equation}
C_R(\omega') = (1/4) \hbox{Tr}\left[(\sigma_y \otimes \sigma_y)^2\right] = 1.  
\end{equation}
That is, in the real-vector-space theory,
the state $\omega'$ is maximally entangled, even though the same density matrix represents an unentangled state in the complex-vector-space theory; that is, $C(\omega') = 0$.  The difference comes from the extra freedom one has in the complex theory, namely, the freedom to use complex states in the decomposition of the density matrix.  This mathematical difference represents a genuine physical distinction between the two theories.  If we lived in a world governed by real-vector-space quantum theory, two spatially separated participants would not be able generate the state $\omega'$, shared between them, without sending a quantum particle over the intervening space (assuming they share no entanglement initially).  This is because in the real-vector-space theory, {\em every} pure-state decomposition of the state $\omega'$ consists of nothing but maximally entangled states.  And yet in our actual, complex-vector-space quantum world, there exists a product-state decomposition and we can therefore create the state $\omega'$ without any quantum transmission.  

We can already see an important difference between the real-vector-space theory and ordinary quantum theory.  In ordinary quantum theory, the only maximally entangled states are pure states.  But in the real-vector-space theory, the state $\omega'$, which is mixed, is nevertheless maximally entangled.  That such a case is possible has been pointed out in Refs.~\cite{Fuchs} and \cite{Batle}.

\section{Sharing entanglement}

Here we show that the real-vector-space theory strongly violates entanglement monogamy.  

We begin by writing down two pure states of three rebits, each of which has the property
that every pair of rebits is maximally entangled (in the real-vector-space sense).  The two states are
\begin{equation}
|\mu_{++}\rangle =  (1/2)(|000\rangle - |011\rangle - |101\rangle - |110\rangle)\label{pure1}
\end{equation}
and
\begin{equation}
|\nu_{++}\rangle = (1/2)(001\rangle + |010\rangle + |100\rangle - |111\rangle)  .  \label{pure2}
\end{equation}
(The meaning of the subscript $++$ will become clear shortly.)  For each of these states, if we trace over 
any one of the three rebits, we find that the state of the other two is given by the density matrix $(1/2)(|\Phi^-\rangle\langle\Phi^-| + |\Psi^+\rangle\langle \Psi^+|)$, which is equal to the state $\omega'$ we analyzed in the preceding
section.  As we have seen, this state has unit concurrence.  Thus each pair of rebits is indeed maximally entangled,
as could never be the case in standard quantum mechanics.

We can generalize the states in Eqs.~(\ref{pure1}) and (\ref{pure2}) to an arbitrary number of rebits.  In order to 
do the generalization, we first rewrite the states as follows.  Let $|y_+\rangle$ and $|y_-\rangle$ be eigenstates
of $\sigma_y$ with eigenvalues $+1$ and $-1$ respectively:
\begin{equation}
|y_+\rangle = \frac{1}{\sqrt{2}}\mtx{c}{1 \\ i}  \hspace{1cm} |y_-\rangle = \frac{1}{\sqrt{2}}\mtx{c}{1 \\ -i} .
\end{equation}
We can express the above states as
\begin{equation}
|\mu_{++}\rangle =\sqrt{2}\, \hbox{Re}\, (|y_+\rangle \otimes |y_+\rangle \otimes |y_+\rangle) \hspace{1cm} 
|\nu_{++}\rangle = \sqrt{2}\, \hbox{Im}\, (|y_+\rangle \otimes |y_+\rangle \otimes |y_+\rangle).
\end{equation}

Guided by the form of these last equations, we now write down, for a system of $n$ rebits, a set of $2^n$ pure states, each of which has the property that every
pair of rebits is maximally entangled.  The states are indexed by binary indices $s_1$, \ldots, $s_{n-1}$, 
each of which takes the values $+1$ and $-1$.
\begin{equation}
|\mu_{s_1\ldots s_{n-1}}\rangle = \sqrt{2}\, \hbox{Re}(|y_{s_1}\rangle \otimes \cdots
\otimes |y_{s_{n-1}}\rangle \otimes |y_+\rangle), \label{re}
\end{equation}
\begin{equation}
|\nu_{s_1\ldots s_{n-1}}\rangle = \sqrt{2}\, \hbox{Im}(|y_{s_1}\rangle \otimes \cdots
\otimes |y_{s_{n-1}}\rangle \otimes |y_+\rangle).  \label{im}
\end{equation}
Note that it would be redundant to include another binary index $s_n$.  In $|\mu_{s_1\ldots s_{n-1}}\rangle$, for example,
if we were to change the sign of each $s_i$, and in addition change the last factor from $|y_+\rangle$ to
$|y_-\rangle$, we would be taking the complex conjugate of the whole state, and the real part would remain unchanged.
Similarly, the imaginary part that appears in Eq.~(\ref{im}) would merely switch its sign under this operation, and an overall sign change does not affect the
physical properties of a state vector. We note also that all $2^n$ states defined in Eqs.~(\ref{re}) and (\ref{im})
are mutually orthogonal: states corresponding to different values of $(s_1,\ldots, s_{n-1})$ are orthogonal because of the
orthogonality of $|y_+\rangle$ and $|y_-\rangle$ (as can be seen from Eq.~(\ref{musum}) below), and for any given value of $(s_1,\ldots, s_{n-1})$, the states
$|\mu\rangle$ and $|\nu\rangle$ are orthogonal because each component of the vector $|y_{s_1}\rangle \otimes \cdots
\otimes |y_{s_{n-1}}\rangle \otimes |y_+\rangle$ (in the standard basis) is either purely real or purely 
imaginary. 

To show that these states have maximal entanglement between every pair of rebits, we compute the
average value of the operator $\sigma_y^{(i)} \otimes \sigma_y^{(j)}$ for any given pair $(i,j)$.  (Here we leave implicit
the $n-2$ factors of the identity matrix associated with the other rebits.)  We begin with 
the state $|\mu_{s_1\ldots s_{n-1}}\rangle$, which we can write as 
\begin{equation}  \label{musum}
|\mu_{s_1\ldots s_{n-1}}\rangle =
(1/\sqrt{2})(|y_{s_1}\rangle \otimes \cdots
\otimes |y_{s_{n-1}}\rangle \otimes |y_+\rangle + |y_{-s_1}\rangle \otimes \cdots
\otimes |y_{-s_{n-1}}\rangle \otimes |y_-\rangle).
\end{equation}
Since $|y_+\rangle$ and $|y_-\rangle$ are eigenstates of $\sigma_y$, it is straightforward to compute the 
average value of $\sigma_y^{(i)} \otimes \sigma_y^{(j)}$.  One finds that
\begin{equation}
 \langle \mu_{s_1\ldots s_{n-1}}|\left( \sigma_y^{(i)} \otimes \sigma_y^{(j)}\right) | \mu_{s_1\ldots s_{n-1}} \rangle= s_is_j,
\end{equation}
where we define $s_n$ to be equal to $+1$.  
A similar calculation shows that $ | \nu_{s_1\ldots s_{n-1}} \rangle $ yields the same average value.  The real-vector-space 
concurrence between particles $i$ and $j$ is the magnitude of the average value of $\sigma_y^{(i)} \otimes \sigma_y^{(j)}$.  
So we see that this concurrence indeed has the value 1 for every pair, as claimed.  

In addition to the pure states considered above, it is interesting to consider a specific set of mixed states, also
labeled by the indices $s_1, \ldots, s_{n-1}$:
\begin{equation}
\rho_{s_1 \ldots s_{n-1}} = (1/2)(|\mu_{s_1 \ldots s_{n-1}}\rangle\langle\mu_{s_1 \ldots s_{n-1}}| + 
|\nu_{s_1 \ldots s_{n-1}}\rangle\langle\nu_{s_1 \ldots s_{n-1}}| ).   \label{mixed}
\end{equation}
Since for any pair $(i,j)$ the two pure states in this expression have the same average value of $\sigma_y^{(i)} \otimes \sigma_y^{(j)}$,
the state $\rho_{s_1 \ldots s_{n-1}}$ shares this value, and therefore, like the pure states of which it is composed,
it also exhibits maximal pairwise entanglement.  

It will be convenient to write the state $\rho_{s_1\ldots s_{n-1}}$ in another form:
\begin{equation}
\rho_{s_1s_2\ldots s_{n-1}} = \bigg(\frac{1}{2^n}\bigg)\hbox{Re}\bigg[\sum_{j_1,\ldots,j_{n}=0}^1
(s_1 \sigma_y)^{j_1} \otimes \cdots \otimes (s_{n-1}\sigma_y)^{j_{n-1}} \otimes   \label{rhos}
\sigma_y^{j_n} \bigg].
\end{equation}
Taking the real part in effect limits the summation to the cases where an even number of the exponents $j_i$
are equal to 1, since $\sigma_y$ is a purely imaginary matrix.  One way to see that Eq.~(\ref{rhos}) is equivalent
to Eq.~(\ref{mixed}) is to start with Eq.~(\ref{rhos}) and explicitly write down all of the eigenstates of the
operator.  The eigenstates can be taken to be all $n$-fold tensor products of $|y_+\rangle$ and $|y_-\rangle$.
One finds that all the eigenvalues are zero except in the case of two eigenvectors, which lie in the subspace spanned 
by $|\mu_{s_1\ldots s_{n-1}}\rangle$ and $|\nu_{s_1\ldots s_{n-1}}\rangle$.  In that subspace, each eigenvalue
is 1/2.  Thus the eigenvalues and eigenvectors of the matrix in Eq.~(\ref{rhos}) are the same as for the 
matrix in Eq.~(\ref{mixed}).  So the two matrices are the same.

We can use the form (\ref{rhos}) to demonstrate a remarkable property of the mixed states $\rho_{s_1\ldots s_{n-1}}$.  Even though, in each state, each pair of rebits
is maximally entangled, there are {\em no classical correlations} among the outcomes of any local measurements.
To see this, suppose that an observer located at the site of each rebit performs some measurement just on that rebit.  Any particular outcome
of a measurement on rebit $i$ is associated with some positive operator $\Pi_i$, and for a particular state $\rho$, the probability of getting 
a specific collection of local outcomes is given by
\begin{equation}
p = \hbox{Tr}[(\Pi_1 \otimes \cdots \otimes \Pi_n) \rho].
\end{equation}
In the real-vector-space theory, each $\Pi_i$ is a real symmetric operator.
But in Eq.~(\ref{rhos}), all but one of the terms in the sum has at least two factors of $\sigma_y$, which is
an antisymmetric matrix.  The trace of the 
product of a symmetric matrix and an antisymmetric matrix is zero.  So the only term that contributes to the 
probability is the identity term ($j_1 = \cdots = j_n = 0$), and we therefore have 
\begin{equation}
p = (1/2^n) \hbox{Tr} (\Pi_1 \otimes \cdots \otimes \Pi_n) = \left( \frac{1}{2} \hbox{Tr}\, \Pi_1\right)\times
\cdots \times \left( \frac{1}{2} \hbox{Tr}\, \Pi_n\right).
\end{equation}
That is, the overall probability distribution is a product of local probability distributions; there are no correlations.
Note also that the probabilities do not depend on the indices $s_1, \ldots, s_{n-1}$.  All of the states
defined in Eq.~(\ref{rhos}) have identical statistics with regard to collections of local measurements.  Essentially,
the outcome of any such collection of measurements is entirely random, as if the system were in the completely mixed state. (But it is not in that state.  In real-vector-space quantum theory, the statistics of the outcomes of local measurements are not sufficient to uniquely determine the global density matrix \cite{tomography}.) 

It may seem a contradiction that a state can exhibit maximal pairwise entanglement and yet show no 
correlations among local measurements.  This combination of properties certainly cannot occur in ordinary
quantum mechanics, but it is not a logical contradiction.  The states $\rho_{s_1 \ldots s_{n-1}}$ are entangled in the sense that
they cannot be created without quantum communication (in a real-vector-space quantum world), but the presence of
entanglement in this sense does not imply anything about correlations among measurements.  (On the other hand,
the pure states $|\mu_{s_1 \ldots s_{n-1}}\rangle$ and $|\nu_{s_1 \ldots s_{n-1}}\rangle$ do exhibit nontrivial
correlations, but only at the $n$-rebit level.  For example, if three rebits are in the state $|\mu_{++}\rangle$
of Eq.~(\ref{pure1}), then there are no correlations between any pair, but if each of the first two rebits gives the outcome ``$0$''
in a ``0 vs 1'' measurement, the third rebit will also yield ``0''.)

Now, suppose we lived in a world governed by real-vector-space quantum theory.  It is interesting to ask whether we
could do anything practical with the above states that makes use of their strange entanglement
properties.  At least with the states $\rho_{s_1 \ldots s_{n-1}}$, there is such an application: we could hide classical data \cite{hiding}.  Someone who wanted to hide data could create one of
these states, with $n-1$ bits of classical data encoded in the indices $s_1, \ldots, s_{n-1}$.  Suppose the $n$ rebits are
placed in different locations, with an observer stationed at each site.  And suppose that these observers are not
able to transmit quantum information from one site to another, because there are no reliable quantum
channels connecting the sites.  The observers can try to obtain information about the 
bits $s_1, \ldots, s_{n-1}$ by making local measurements and comparing the outcomes, but as we have just seen, they will not succeed.  The probabilities of the outcomes of local measurements are completely independent of 
the values of $s_1, \ldots, s_{n-1}$.  So the hidden data is safe as long as quantum communication among the 
sites is impossible.  On the other hand, if all the observers were to bring their rebits to a common location,
they would be able to completely determine all $n-1$ bits: they could distinguish the $2^{n-1}$ states
from each other since the states are orthogonal.  

\section{Discussion}

We have seen that there is dramatic difference between ordinary quantum theory and the real-vector-space theory with regard
to the sharing of entanglement.  
In the former theory, entanglement is monogamous, whereas in the latter, there are multiparticle states exhibiting maximal 
entanglement between any two particles.  It is important to reiterate that we are defining entanglement based on
a consideration of the resources needed to create a given state, not with regard to correlation or to the possibility of distilling pure entanglement.  It is conceivable that for other notions of entanglement, some form of entanglement monogamy may be retained in the real-vector-space theory.  

Some authors have noted that a system of $n$ qubits, following the rules of ordinary quantum theory, can be simulated by a system of $n+1$ rebits \cite{simulation1,simulation2,simulation3}.  (The last of these papers shows how, with additional rebits, one can make the simulation local.)  
The extra rebit has the effect of doubling the dimension of the state space, and a $2^{n+1}$-dimensional real vector
space is large enough to accommodate all of the structure of a $2^n$-dimensional complex vector space.  (In a sense it
is too large: one has to restrict the set of allowed observables and transformations in order to recover standard
quantum theory \cite{Stueckelberg}.)  This observation is related to the work presented in this paper.  In order for a single extra rebit to 
``complexify'' the state space of each of the other $n$ rebits, the extra rebit must be able to be entangled with
each of the other rebits simultaneously.  If entanglement monogamy were in force in the real-vector-space theory, such simultaneous entanglement would not be possible.  Thus the fact that the real-vector-space theory can simulate the 
complex theory, with only a single added binary object, already suggests that the real-vector-space theory does not respect entanglement monogamy in the usual sense.  

We noted in the Introduction that in classical probability theory there is no limitation on the sharing of correlations.  And we have seen that for a system of rebits there is no limitation on the sharing of entanglement as measured by concurrence.  This odd similarity between classical probability theory and real-vector-space quantum theory actually extends a little further.  Let us ask how many states (that is, how many probability distributions) of $n$ classical coins exhibit maximal correlation between any two coins.  The answer is $2^{n-1}$.  Given any sequence of $n$ heads and tails, we can construct one of these probability distributions by assigning probability 1/2 to the chosen sequence and probability 1/2 to the ``opposite'' sequence in which heads and tails are interchanged.  
There are $2^n$ sequences, but any sequence and its opposite give rise to the same probability distribution in this construction; so there are only $2^{n-1}$ distinct probability distributions having 
this property.  There is a sense in which these $2^{n-1}$ classical states are the mirror image of the $2^{n-1}$ mixed 
states $\rho_{s_1\ldots s_{n-1}}$.  Each of the special states of the classical theory has zero entanglement and maximal pairwise correlation.  In contrast, each special state $\rho_{s_1\ldots s_{n-1}}$ has maximal pairwise entanglement and no correlations.  

Quantum theory occupies an interesting middle ground between these two extremes.  Entanglement is possible but limited, and pairwise entanglement always entails pairwise correlation.  Perhaps the main lesson of what we have done here is that the usual limitation on entanglement sharing is more specific to standard quantum theory than one might have thought, in that it is lost when we remove the possibility of using ``$i$'' in our state vectors.

\end{document}